\documentclass[12pt]{article}
\usepackage{epsf}
\def\beq{\begin{equation}}
\def\eeq{\end{equation}}
\def\ap#1#2#3 {Ann. Phys. (NY) {\bf#1} (19#2) #3}
\def\err#1#2#3 {{\it Erratum} {\bf#1} (19#2) #3}
\def\ib#1#2#3 {{\it ibid.} {\bf#1} (19#2) #3}
\def\ijmp#1#2#3 {Int. J. Mod. Phys. {\bf#1} (19#2) #3}
\def\jetp#1#2#3 {JETP Lett. {\bf#1} (19#2) #3}
\def\mpl#1#2#3 {Mod. Phys. Lett. {\bf#1} (19#2) #3}
\def\np#1#2#3 {Nucl. Phys. {\bf#1} (19#2) #3}
\def\pl#1#2#3 {Phys. Lett. {\bf#1} (19#2) #3}
\def\prep#1#2#3 {Phys. Rep. {\bf#1} (19#2) #3}
\def\prev#1#2#3 {Phys. Rev. {\bf#1} (19#2) #3}
\def\prl#1#2#3 {Phys. Rev. Lett. {\bf#1} (19#2) #3}
\def\sjnp#1#2#3 {Sov. J. Nucl. Phys. {\bf#1} (19#2) #3}
\def\spj#1#2#3 {Sov. Phys. JETP {\bf#1} (19#2) #3}
\def\spu#1#2#3 {Sov. Phys. Usp. {\bf#1} (19#2) #3}
\def\zp#1#2#3 {Zeit. Phys. {\bf#1} (19#2) #3}

\begin{document}
\begin{titlepage}
\begin{center}
{\Large \bf Theoretical Physics Institute \\
University of Minnesota \\}  \end{center}
\vspace{0.2in}
\begin{flushright}
TPI-MINN-99/40-T \\
UMN-TH-1815-99 \\
hep-th/9908455\\
August 1999 \\
\end{flushright}
\vspace{0.3in}
\begin{center}
{\Large \bf  Reducing model dependence of spectator effects in inclusive
decays of heavy baryons
\\}
\vspace{0.2in}
{\bf M.B. Voloshin  \\ }
Theoretical Physics Institute, University of Minnesota, Minneapolis,
MN
55455 \\ and \\
Institute of Theoretical and Experimental Physics, Moscow, 117259
\\[0.2in]
\end{center}

\begin{abstract}

The dependence of inclusive weak decay rates of heavy hadrons on light
spectator quarks is considered. The analysis of a previous work on
relating the effects in $b$ baryons to those in charmed baryons is
extended to explicit evaluation of the matrix elements of certain
four-quark operators over heavy baryons. It is shown that the usually
postulated color antisymmetry of these matrix elements is significantly
broken. The flavor-singlet shift of inclusive decay rates of $b$ baryons
due to the spectator effects is shown to be strongly suppressed in the
leading-log approximation. Combined with the results for the flavor
non-singlet splittings, this observation allows to argue, in a less
model-dependent way, than before, that within the present understanding
of the spectator effects it is highly unlikely that the lifetimes of the
$\Lambda_b$ and the $B_d$ can be split by more than 10\%.

\end{abstract}
\end{titlepage}

\section{Introduction}

The differences in inclusive weak decay rates of hadrons, containing one
heavy quark, are known \cite{pdg} to be very prominent for the charmed
mesons and baryons and, although are substantially smaller for the $b$
hadrons, are well measurable and present a subject for an interesting
experimental and theoretical study. Theoretically these differences are
understood in the framework of the operator product expansion {\rm
[2-5]} in inverse powers of the heavy quark mass $m_Q$ (for a later
review see e.g. Ref. \cite{bs}). The leading term in this expansion is
the `parton' decay rate of the heavy quark $\Gamma_{part} \propto m_Q^5$
and does not depend on the light `environment' surrounding the heavy
quark $Q$ in a hadron. Thus this term, setting the overall scale for the
decay rates of hadrons containing the heavy quark $Q$, provides no
splitting between the decay widths of specific hadrons. The first
subleading term, suppressed with respect to the leading one by
$m_Q^{-2}$, describes the effects of motion of the heavy quark in a
hadron (time dilation) and the effects of the chromomagnetic interaction
of the heavy quark \cite{buv}. This term distinguishes between mesons
and baryons and between baryons of different spin structure, but
provides no splitting of the decay rates within flavor SU(3) multiplets
of heavy hadrons. The dependence on the spectator (anti)quark flavor
arises in the next order, $m_Q^{-3}$ relative to $\Gamma_{part}$, and is
expressed through matrix elements of local four-quark operators over
particular heavy hadron $X_Q$: $\langle X_Q | ({\overline Q}\, \Gamma \,
Q) \, ({\overline q} \, \Gamma^{'} \, q) | X_Q \rangle$ with different
light quark flavors $q$ and with different spin-color matrix structures
$\Gamma$ and $\Gamma^{'}\,$\footnote{The physical origin of these terms
is traditionally explained as due to the Pauli interference (PI) and the
weak annihilation (WA){\rm [2-6]}. However in practical use of the OPE
this terminology is somewhat redundant. Moreover, the physical
interpretation of a particular four-quark operator as PI or WA depends
on the hadron discussed.}. It should be noted that although formally of
higher order in $m_Q^{-1}$, the effects of the four-quark operators are
generally larger than those of the $O(m_Q^{-2})$ terms for charmed
hadrons and are larger than or comparable to the latter in the $b$
hadrons, due to a large numerical factor.

The principal obstacle for quantitatively describing the
spectator-dependent differences in the decay rates of heavy hadrons
within this OPE-based theoretical scheme is our ignorance about the
hadronic matrix elements of the relevant four-quark operators. For
mesons one usually appeals to factorization and approximately expresses
these matrix elements in terms of the mesons' annihilation constants $f$
{\rm [2-6]}, with the thus arising uncertainty being relegated to the
``bag constant" $B$. For the baryons the situation is even worse: most
of the estimates operate with a naive (constituent, valence) quark
model, where these matrix elements are expressed in terms of ``wave
function at the origin" $|\psi(0)|^2$, and then it is either assumed
that this quantity is the same as in mesons, or arguments for its
suppression or enhancement in baryons are presented (see e.g. \cite{bgt,
grt, rosner, ngu}). Although this simplistic approach proved to be
successful in qualitatively predicting the hierarchy of the lifetimes of
charmed hadrons \cite{sv3}, its accuracy is clearly insufficient for a
better quantitative description of the differences in the inclusive
decay rates.

One of the problems, currently attracting a considerable interest, is
presented by the measured difference of the lifetimes of the $B$ mesons
and of the $\Lambda_b$ baryon \cite{pdg}:
$\tau(\Lambda_b)/\tau(B_d)=0.81 \pm 0.05$. The effect of the
$O(m_b^{-2})$ terms in this splitting is believed to be less than  2\%
\cite{bbsuv, ns}. Thus, if confirmed by further measurements, this large
difference in the lifetimes will have to be explained either by the
contribution of the four-quark operators, or by effects beyond the
OPE-based approach. So far all calculations based on the quark model
approach as well as a calculation using QCD sum rules \cite{cf} have
failed to find an enhancement of the four-quark matrix elements
sufficient to achieve a 20\% enhancement of the decay rate of
$\Lambda_b$, required by the current data. The largest predictions for
this enhancement go to only about 10\%.

In view of this discrepancy between the available data and the
calculations, it is highly desirable to reduce the model
dependence of the theoretical predictions for the relevant four-quark
matrix elements over heavy hadrons. One approach in this direction is to
use available experimental data to evaluate the required matrix elements
and then to apply the results to predict other differences in
inclusive decay rates {\cite{mv1, mv2}. In particular, in this way one
can relate the differences in decay rates of the $b$ hadrons to those
for the charmed hadrons. Certainly, such approach is also not
fully proof. One possible source of uncertainty is the application of
the heavy quark expansion to charmed hadrons. The mass of charmed quark
is by no means asymptotically heavy, thus subsequent terms in $m_c^{-1}$
potentially can be large. One of obvious manifestations of a
relatively low $m_c$ is that the spectator effects in charmed hadrons
are in fact dominating in the decay rates, rather than being small
corrections. This however does not necessarily invalidate the
applicability of the discussed OPE expansion to charmed hadrons, since
the enhancement of the $m_Q^{-3}$ terms by a large numerical factor has
no recurrence in subsequent terms of the expansion. Another uncertainty
arises from using the flavor SU(3) symmetry in order to relate matrix
elements within one SU(3) multiplet. Both approximations, however can be
tested by comparing the predictions within this approach \cite{mv2} for
inclusive decay rates into various channels (semileptonic and CKM
suppressed), once the corresponding experimental data become available.

The purpose of the present paper is to extend the analysis of Ref.
\cite{mv2} of the decay rate splittings in the triplet of heavy baryons
($\Lambda_Q$ and two $\Xi_Q$'s) to discussion of particular four-quark
matrix elements over heavy baryons, and to an analysis of the overall
(flavor SU(3)-singlet) shift of the inclusive decay rates for the
baryonic triplet in addition to the splittings within the triplet.

As will be demonstrated here, although the data on the lifetimes of
charmed baryons imply a significant enhancement (by a factor of 4-6) of
the four-quark matrix elements, the color structure comes out entirely
different from the expectations based on naive quark model.
Namely, inherent in an naive quark model approach is the color
antisymmetry relation for the baryonic matrix elements:
\beq
\langle X_Q | ({\overline Q}_i\, \Gamma \, Q^i) \, ({\overline q}_j \,
\Gamma^{'} \, q^j) | X_Q \rangle =- \langle X_Q | ({\overline Q}_i\,
\Gamma \, Q^j) \, ({\overline q}_j \, \Gamma^{'} \, q^i) | X_Q \rangle,
\label{cas}
\eeq
where this time $\Gamma$ and $\Gamma^{'}$ stand for colorless spinor
matrices. The color antisymmetry also holds in the sum rules calculation
\cite{cf} within the level of complexity adopted there. Also this
relation was used as an input in the most recent paper \cite{gms} on the
subject.
The relation (\ref{cas}) however can not be valid at all normalization
scales $\mu$ for the operators due to different $\mu$ dependence of the
color structures involved. Thus the validity and the meaning of the
color antisymmetry are questionable at the least. It will be shown here
that the data imply that the matrix element in the left-hand-side of
eq.(\ref{cas}) is significantly enhanced, while that in the
right-hand-side is rather suppressed and is likely to be negligible or
zero at a low normalization point $\mu$ such that $\alpha_s(\mu) \sim
1$.

The enhancement of certain matrix elements, derived from the data on
decays of charmed baryons, might inspire enthusiasm in explaining the
data on $\tau(\Lambda_b)/\tau(B_d)$. Indeed, the result of Ref.
\cite{mv2} for the splitting of the total decay rates between
$\Lambda_b$ and $\Xi_b^-$: $\Gamma (\Lambda_b) - \Gamma (\Xi_b^-) = 0.11
\pm 0.03 \, ps^{-1}$, which constitutes $(14 \pm 4)$\% of the measured
width of $\Lambda_b$, might hint that shifts of the rates by the
spectator effects by about 10 or 20\% should not be uncommon among the
$b$ baryons\footnote{Another very simplistic reasoning can be as
follows. The spectator effects in the lifetimes of charmed baryons
amount to few $ps^{-1}$. When rescaled to the $b$ baryons by the factor
$|V_{cb}|^2 \, m_c^2/m_b^2 \approx 0.015$, this puts the ``natural"
magnitude of these effects in the ballpark of $0.1 \, ps^{-1}$.}.
However, as will be discussed, the splitting $\Gamma (\Lambda_b) -
\Gamma (\Xi_b^-)$ turns out to be the largest spectator effect for the
triplet of $b$ baryons. The inclusive decay rates of $\Lambda_b$ and
$\Xi_b^0$ are degenerate up to terms additionally suppressed by
$m_c^2/m_b^2$, while the overall (SU(3)-singlet) shift of the decay
rates is greatly suppressed by a small coefficient in the flavor-singlet
part of the effective Lagrangian describing the spectator effects.
Parametrically, in the leading-log order (LLO), the suppression factor
is proportional to $[(\alpha_s/\pi) \, \ln m_W/m_b]^2$, and (with
appropriate numerical factors) amounts to about 0.1. Thus the average
shift of the decay rate in the baryon triplet is expected to be only of
the order of $10^{-2}\, ps^{-1}$. When combined with the estimate of the
chromomagnetic $O(m_b^{-2})$ effects and the prediction for the
splitting of the rates within the $b$ baryon triplet, the central value
of the expected enhancement of the total decay width of $\Lambda_b$ as
compared to $B_d$ does not exceed 8\%. In view of the suppression of the
discussed SU(3) singlet shift in the LLO approximation, the
next-to-leading-log (NLLO) terms can in fact be as much essential as the
LLO ones. However, it is highly unlikely that they can exceed the latter
by a factor of more than 10. Thus we conclude that within an analysis,
not dealing with a model of quark wave functions of heavy baryons, the
central value of the current data on $\tau(\Lambda_b)/\tau(B_d)$ can not
be accommodated in the framework of the OPE-based description of
inclusive decay rates of heavy hadrons.

The rest of the paper is organized as follows. In Section 2 the OPE
approach to spectator effects in inclusive decays of heavy hadrons is
briefly described and the relevant to the reasoning in this paper parts
of the effective Lagrangian with four-quark operators are presented. In
Section 3 these expressions are used, continuing along the lines of Ref.
\cite{mv2}, to extract certain combinations of the four-quark matrix
elements over heavy baryons. In Section 4 the suppression in the
leading-log order of the flavor-singlet shift of the decay rates of the
(anti)triplet of the $b$ baryons is demonstrated, and possible effects
beyond that order are discussed in Section 5.

\section{Spectator dependent terms in the OPE approach}

The description of the leading in the limit $m_Q \to \infty$ as well as
subleading effects in
inclusive decay rates of heavy hadrons arises through application of
operator product expansion to the `effective Lagrangian' related to the
correlator of the weak-interaction Lagrangian $L_W$:
\beq
L_{eff}=2 \,{\rm Im} \, \left [ i \int d^4x \, e^{iqx} \, T \left \{
L_W(x),
L_W(0) \right \} \right ]~.
\label{leff}
\eeq
The inclusive decay rate of a heavy hadron $X_Q$ is then determined as
\beq
\Gamma_X=\langle X_Q | \, L_{eff} \, | X_Q \rangle~,
\label{lgam}
\eeq
within the adopted throughout this paper non-relativistic normalization
for the heavy quark states: $\langle Q | Q^\dagger Q | Q \rangle =1$.
The subject of interest in this paper, the dependence on the spectator
quarks, is described by the term in the OPE for $L_{eff}$, containing
the light quark fields. This term is denoted here as $L_{eff}^{(3)}$,
indicating that it is the third term in the operator expansion (after
the leading one and those of order $m_Q^{-2})$. The expressions for the
parts of $L_{eff}^{(3)}$ describing inclusive decay rates of the charmed
and $b$ hadrons into specific flavor-types of final states are found by
picking the corresponding parts of the weak Lagrangian $L_W$. For the
dominant unsuppressed by the CKM mixing nonleptonic decays of charmed
hadrons, associated with the underlying quark process $c \to s \, u \,
{\overline d}$, the relevant part of $L_{eff}^{(3)}$ reads as \cite{sv3}
\begin{eqnarray}
\label{l3nl}
&&L_{eff, \, nl}^{(3, 0)}= \cos^4 \theta_c \,{G_F^2 \, m_c^2 \over 4
\pi} \,
\left \{
C_1 \, (\overline c \Gamma_\mu c)(\overline d \Gamma_\mu d) + C_2  \,
(\overline c \Gamma_\mu d) (\overline d \Gamma_\mu c) +\right .
\nonumber \\
&& C_3 \, (\overline  c \Gamma_\mu c +
{2 \over 3}\overline c \gamma_\mu \gamma_5 c) (\overline s \Gamma_\mu
s)+ C_4 \, (\overline  c_i \Gamma_\mu c_k +
{2 \over 3}\overline c_i \gamma_\mu \gamma_5 c_k)
(\overline s_k \Gamma_\mu s_i) +
\\ \nonumber
&& C_5 \, (\overline  c \Gamma_\mu c +
{2 \over 3}\overline c \gamma_\mu \gamma_5 c) (\overline u \Gamma_\mu
u)+ C_6 \, (\overline  c_i \Gamma_\mu c_k +
{2 \over 3}\overline c_i \gamma_\mu \gamma_5 c_k)
(\overline u_k \Gamma_\mu u_i)+ \\ \nonumber
&&\left. {1 \over 3} \,
\kappa^{1/2} \, (\kappa^{-2/9}-1) \, \left [ 2 \, (C_+^2 - C_-^2) \,
 (\overline c \Gamma_\mu t^a c) \,
j_\mu^a - (5C_+^2+C_-^2)
(\overline  c \Gamma_\mu t^a c +
{2 \over 3}\overline c \gamma_\mu \gamma_5 t^a c) j_\mu^a \right ]
\right \}~
\end{eqnarray}
where $i, \, k$ are color indices, $\Gamma_\mu=\gamma_\mu \,
(1-\gamma_5)$, and the coefficients $C_A, ~A=1, \ldots, 6$ are given by
\begin{eqnarray}
&&C_1= C_+^2+C_-^2 + {1 \over 3} (1 - \kappa^{1/2}) (C_+^2-C_-^2)~,
\nonumber \\
&&C_2= \kappa^{1/2} \, (C_+^2-C_-^2)~, \nonumber \\
&&C_3=- {1 \over 4} \, \left [ (C_+-C_-)^2 + {1 \over 3}
(1-\kappa^{1/2})
(5C_+^2+C_-^2+6C_+C_-) \right] ~, \nonumber \\
&&C_4=-{1 \over 4} \, \kappa^{1/2} \, (5C_+^2+C_-^2+6C_+C_-)~, \nonumber
\\
&&C_5=-{1 \over 4} \, \left [ (C_++C_-)^2 + {1 \over 3} (1-\kappa^{1/2})
(5C_+^2+C_-^2-6C_+C_-) \right]~, \nonumber \\
&&C_6=-{1 \over 4} \, \kappa^{1/2} \, (5C_+^2+C_-^2-6C_+C_-)~.
\label{coefs}
\end{eqnarray}
in terms of the well know short-distance renormalization factors $C_+$
and $C_-$ for the nonleptonic weak interaction:
$C_-=C_+^{-2}=(\alpha_s(m_c)/\alpha_s(m_W))^{4/b}$ with $b$ being the
coefficient in the one-loop beta function in QCD. For the case of the
charmed quark decay one can use $b=25/3$ (see e.g. in the textbook
\cite{lbo}). Furthermore, the parameter
$\kappa=(\alpha_s(\mu)/\alpha_s(m_c))$ describes the so-called `hybrid'
renormalization of the operators below the heavy quark mass and down to
a low normalization point $\mu$, and $j_\mu^a=\overline u \gamma_\mu t^a
u + \overline d \gamma_\mu t^a d +
\overline s \gamma_\mu t^a s$ is the standard notation for the color
current of light quarks with $t^a=\lambda^a/2$ being the generators of
the color SU(3) group.

Besides the dominant nonleptonic decays, it is expected \cite{mv1, mv2}
that the spectator effects in the once CKM suppressed nonleptonic and
the semileptonic inclusive decay rates of charmed hyperons are quite
large, and produce a non-negligible impact on the lifetimes of these
hyperons. Therefore in order to analyze the four-quark matrix elements
from the data on the lifetimes of the charmed baryons, one has to
consider also the corresponding terms in $L_{eff}^{(3)}$. The once CKM
suppressed part, associated with the quark processes $c \to s \, u \,
{\overline s}$ and $c \to d \, u \, {\overline d}$ has the form
\cite{mv2}
\begin{eqnarray}
\label{l3nl1}
&&L_{eff, \, nl}^{(3, 1)}= \cos^2 \theta_c \, \sin^2 \theta_c \,{G_F^2
\, m_c^2 \over 4 \pi} \,
\left \{
C_1 \, (\overline c \Gamma_\mu c)(\overline q \Gamma_\mu q) + C_2  \,
(\overline c_i \Gamma_\mu c_k) (\overline q_k \Gamma_\mu q_i) +\right .
\\ \nonumber
&& C_3 \, (\overline  c \Gamma_\mu c +
{2 \over 3}\overline c \gamma_\mu \gamma_5 c) (\overline q \Gamma_\mu
q)+ C_4 \, (\overline  c_i \Gamma_\mu c_k +
{2 \over 3}\overline c_i \gamma_\mu \gamma_5 c_k)
(\overline q_k \Gamma_\mu q_i) +  \\ \nonumber
&& 2 \, C_5 \, (\overline  c \Gamma_\mu c +
{2 \over 3}\overline c \gamma_\mu \gamma_5 c) (\overline u \Gamma_\mu
u)+ 2 \, C_6 \, (\overline  c_i \Gamma_\mu c_k +
{2 \over 3}\overline c_i \gamma_\mu \gamma_5 c_k)
(\overline u_k \Gamma_\mu u_i)+ \\ \nonumber
&&\left. {2 \over 3} \,
\kappa^{1/2} \, (\kappa^{-2/9}-1) \, \left [ 2 \, (C_+^2 - C_-^2) \,
 (\overline c \Gamma_\mu t^a c) \,
j_\mu^a - (5C_+^2+C_-^2)
(\overline  c \Gamma_\mu t^a c +
{2 \over 3}\overline c \gamma_\mu \gamma_5 t^a c) j_\mu^a \right ]
\right \}~
\end{eqnarray}
with the notation $(\overline q \, \Gamma \, q)= (\overline d \, \Gamma
\, d) + (\overline s \, \Gamma \, s)$. The semileptonic part (per one
lepton flavor), generated by the quark decays $c \to s \, \ell^+ \, \nu$
and  $c \to d \, \ell^+ \, \nu$ is given by \cite{mv1, cheng, gm, mv2}
\begin{eqnarray}
&&L_{eff, \, sl}^{(3)}= \nonumber \\
&&{G_F^2 \, m_c^2 \over 12 \pi} \, \left \{  \cos^2 \theta_c \, \left [
L_1 \, (\overline  c \Gamma_\mu c +
{2 \over 3}\overline c \gamma_\mu \gamma_5 c) (\overline s \Gamma_\mu
s)+ L_2 \, (\overline  c_i \Gamma_\mu c_k +  {2 \over 3}\overline c_i
\gamma_\mu \gamma_5 c_k)
(\overline s_k \Gamma_\mu s_i) \right ] + \right. \nonumber \\
&& \sin^2 \theta_c \, \left [
L_1\, (\overline  c \Gamma_\mu c +
{2 \over 3}\overline c \gamma_\mu \gamma_5 c) (\overline d \Gamma_\mu
d)+L_2 \, (\overline  c_i \Gamma_\mu c_k +
{2 \over 3}\overline c_i \gamma_\mu \gamma_5 c_k)
(\overline d_k \Gamma_\mu d_i) \right ] -  \nonumber \\
&& \left. 2 \, \kappa^{1/2} \, (\kappa^{-2/9}-1) \,
(\overline  c \Gamma_\mu t^a c +
{2 \over 3}\overline c \gamma_\mu \gamma_5 t^a c) j_\mu^a
\right \} ~,
\label{l3sl}
\end{eqnarray}
with the coefficients $L_1$ and $L_2$ found as
\beq
L_1=(\kappa^{1/2}-1), ~~~~~
L_2 = -   3\, \kappa^{1/2}~.
\label{coefl}
\eeq

In order to relate the spectator effects in the $b$ hadrons to those in
the charmed ones, one also needs the expression for the part of
$L_{eff}^{(3)}$ describing the $b$ decays. It is sufficient at
the present level of accuracy to retain only the term for the dominant
nonleptonic decays, generated by the quark processes $b \to c \,
{\overline u} \, q$ and $b \to c \, {\overline c}\, q$, where $q$ stands
for $d$ or $s$. This part of $L_{eff}^{(3)}$ is given by \cite{sv3}
\begin{eqnarray}
&&L_{eff, \, nl}^{(3, b)}=  |V_{bc}|^2 \,{G_F^2 \, m_b^2 \over 4
\pi} \,
\left \{
{\tilde C}_1 \, (\overline b \Gamma_\mu b)(\overline u \Gamma_\mu u) +
{\tilde C}_2  \,
(\overline b \Gamma_\mu u) (\overline u \Gamma_\mu b) +\right .
\nonumber \\
&& {\tilde C}_5 \, (\overline  b \Gamma_\mu b +
{2 \over 3}\overline b \gamma_\mu \gamma_5 b) (\overline q \Gamma_\mu
q)+ {\tilde C}_6 \, (\overline  b_i \Gamma_\mu b_k +
{2 \over 3}\overline b_i \gamma_\mu \gamma_5 b_k)
(\overline q_k \Gamma_\mu q_i)+  \nonumber \\
&&\left. {1 \over 3} \,
{\tilde \kappa}^{1/2} \, ({\tilde \kappa}^{-2/9}-1) \, \left [ 2 \,
({\tilde C}_+^2 - {\tilde C}_-^2) \,
 (\overline b \Gamma_\mu t^a b) \,
j_\mu^a - \right. \right. \nonumber \\
&& \left . \left.
(5{\tilde C}_+^2+{\tilde C}_-^2 - 6 \, {\tilde C}_+ \, {\tilde C}_-)
(\overline  b \Gamma_\mu t^a b +
{2 \over 3}\overline b \gamma_\mu \gamma_5 t^a b) j_\mu^a \right ]
\right \}~,
\label{l3nlb}
\end{eqnarray}
where again the notation $(\overline q \, \Gamma \, q)= (\overline d \,
\Gamma \, d) + (\overline s \, \Gamma \, s)$ is used, and the `tilde'
over the renormalization coefficients denotes that these are
determined as described above, albeit with $\alpha_s(m_b)$ instead
of $\alpha_s(m_c)$. The symmetry between $s$ and $d$ quarks ($U$-spin
symmetry) of the expression (\ref{l3nlb}) is a result of the
approximation, where a small kinematical difference of order
$m_c^2/m_b^2$ between the two-body phase space of the quark pairs $c
{\overline c}$ and $c {\overline u}$ is neglected. (The formula with the
full kinematical factors can be found e.g. in \cite{ns}.)

\section{Flavor-nonsinglet matrix elements from differences of lifetimes
of charmed baryons}

When applied to inclusive decay rates of charmed baryons in the SU(3)
(anti)triplet: $(\Lambda_c, \, \Xi_c^+,$ $ \Xi_c^0)$, the expressions in
equations (\ref{l3nl} - \ref{l3sl}) allow one to
extract  matrix elements of certain flavor-nonsinglet four-quark
operators from the data on the differences of lifetimes \cite{mv2}.
Indeed, using the property that there is no correlation of the spin of
the heavy quark with its light `environment' in these baryons, one finds
that the operators with the axial current of the heavy quark do not
contribute to the decay rates of the baryons, so that only the
structures with the vector currents are relevant. These structures are
of the type $(\overline c \, \gamma_\mu \, c)
(\overline q \, \gamma_\mu \, q)$ and $(\overline c_i \, \gamma_\mu \,
c_k)
(\overline q_k \, \gamma_\mu \, q_i)$ with $q$ being $d$, $s$ or $u$.
The difference among the baryons in the triplet of the inclusive decay
rates: the dominant nonleptonic, the once CKM suppressed, and the
semileptonic ones are then all determined in terms of two combinations
of the matrix elements defined as \cite{mv2}
\begin{eqnarray}
\label{defxy}
&&x=\left \langle  {1 \over 2} \, (\overline c \, \gamma_\mu \, c) \left
[
(\overline u \, \gamma_\mu u) - (\overline s \, \gamma_\mu s) \right]
\right \rangle_{\Xi_c^0-\Lambda_c} = \left \langle  {1 \over 2} \,
(\overline c \, \gamma_\mu \, c) \left [ (\overline s \, \gamma_\mu s) -
(\overline d \, \gamma_\mu d) \right] \right \rangle_{\Lambda_c -
\Xi_c^+}~,  \\ \nonumber
&&y=\left \langle  {1 \over 2} \, (\overline c_i \, \gamma_\mu \, c_k)
\left [ (\overline u_k \, \gamma_\mu u_i) - (\overline s_k \, \gamma_\mu
s_i) \right ] \right \rangle_{\Xi_c^0-\Lambda_c} = \left \langle  {1
\over 2} \, (\overline c_i \, \gamma_\mu \, c_k) \left [ (\overline s_k
\, \gamma_\mu s_i) - (\overline d_k \, \gamma_\mu d_i) \right] \right
\rangle_{\Lambda_c - \Xi_c^+}
\end{eqnarray}
with the notation for the differences of the matrix elements:
$\langle {\cal O} \rangle_{A-B}= \langle A | {\cal O} | A \rangle -
\langle B | {\cal O} | B \rangle$.

The formulas for the differences within the charmed baryon triplet of
inclusive rates for individual types of decay can be found in Ref.
\cite{mv2}. Here we make use of only the expressions for the differences
of the total decay rates,
$\Delta_1 \equiv \Gamma(\Xi_c^0)-\Gamma(\Lambda_c)$ and $\Delta_2 \equiv
\Gamma(\Lambda_c)-\Gamma(\Xi_c^+)$, in terms of $x$ and $y$:
\begin{eqnarray}
\Delta_1&=&{G_F^2 \, m_c^2 \over 4 \pi} \, \cos^2 \theta \, \left \{ x
\, \left [ \cos^2 \theta \, (C_5-C_3) + \sin^2 \theta \, (2 \, C_5 - C_1
-C_3)- {2 \over 3} \, L_1 \right ] + \right.
\nonumber \\
&&\left. y \, \left[ \cos^2 \theta \,
(C_6-C_4) + \sin^2 \theta \, (2 \, C_6 - C_2 -C_4)- {2 \over 3} \,
L_2\right ] \right \}~, \nonumber \\
\Delta_2 &=& {G_F^2 \, m_c^2 \over 4 \pi} \ \left \{ x \left[ \cos^4
\theta \, (C_3-C_1) + {2 \over 3} \, (\cos^2 \theta - \sin^2 \theta) \,
L_1 \right] +
\right.
\nonumber \\
&&\left. y \left[ cos^4 \theta \, (C_4-C_2) + {2 \over 3} \,
(\cos^2 \theta - \sin^2 \theta) \, L_2 \right] \right \}~ .
\label{diftot}
\end{eqnarray}

From these formulas one can express the matrix elements $x$ and $y$ in
terms of the data on the lifetimes of the charmed hyperons. In doing so
the values used here are $\Gamma(\Lambda_c)=4.85 \pm 0.28 \, ps^{-1}$,
$\Gamma(\Xi_c^0)= 10.2 \pm 2 \, ps^{-1}$, and the updated value
\cite{pdg1} $\Gamma(\Xi_c^+)=3.0 \pm
0.45 \, ps^{-1}$. For evaluating the short-distance coefficients $C_+$
and $C_-$ a realistic value $\alpha_s(m_c)/\alpha_s(m_W) = 2.5$ is used,
and the results for the matrix elements only weakly depend on
fine-tuning this ratio. As a result the $\mu$ independent matrix element
$x$ is found as
\beq
x = -(0.04 \pm 0.01) \, GeV^3 \, \left ( 1.4 \, GeV \over m_c \right
)^2~,
\label{resx}
\eeq
while the dependence of the thus extracted matrix element $y$ on the
normalization point $\mu$ is shown in Fig.1 \,\footnote{It should be
noted that the curves at large values of $\kappa$, $\kappa > \, \sim 3$,
are shown only for illustrative purpose. The coefficients in the OPE,
leading to the equations (\ref{diftot}), are purely perturbative. Thus,
formally, they correspond to $\alpha_s(\mu) \ll 1$, i.e. to $\kappa \ll
1/\alpha_s(m_c) \sim (3 - 4)$.} .

\begin{figure}[ht]
  \begin{center}
    \leavevmode
    \epsfbox{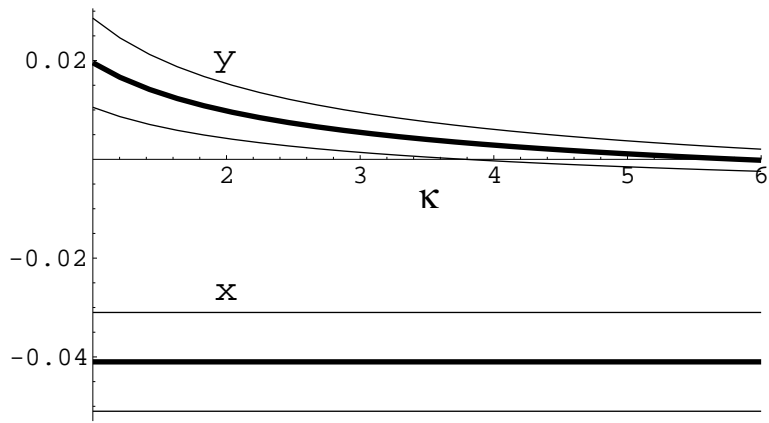}
    \caption{The values of the extracted matrix elements $x$ and $y$ in
$GeV^3$ vs. the normalization point parameter
$\kappa=\alpha_s(\mu)/\alpha_s(m_c)$. The thick lines correspond to the
central value of the data on lifetimes of charmed baryons, and the thin
lines show the error corridors. The extracted values of $x$ and $y$
scale as $m_c^{-2}$ with the assumed mass of the charmed quark, and the
plots are shown for $m_c=1.4 \, GeV$.}
  \end{center}
\label{fig:xy}
\end{figure}

One can see that the extracted values of the matrix elements are in a
drastic variance with naive models. Most conspicuously, the color
antisymmetry relation $x=-y$ fails at all $\mu$ less than $m_c$.
Moreover, the trend in the plot of Fig.1 shows, that a possibly better
approximation would be to assume $y \approx 0$ at a low scale $\mu$
close to the confinement scale. The implication of the vanishing of $y$
is that at large distances there is no correlation between the color of
an individual light quark and the heavy one, which looks quite natural,
if one takes into account strong chaotization of the relative quark
color due to emission of soft gluons. Furthermore, the absolute value of
$x$ significantly exceeds the simplistic estimate \cite{sv2, sv3},
$f_D^2 \, m_D /12 \sim 0.006 \, GeV^3$ of both $|x|$ and $y$ in terms of
the "wave function at the origin" in the baryons related to that in the
mesons. In terms of the coefficient $r$ introduced in \cite{ns} the
result in eq.(\ref{resx}) for $x$ corresponds to a quite large value $r
\approx 5 \pm 1.5$.

\section{Shifts of inclusive decay rates of $b$ baryons}

Once extracted from the data on the lifetimes of the charmed baryons,
the matrix elements can be used for predicting the splitting of
inclusive decay rates of the same baryons into subleading channels and
also for predicting the splittings of the decay rates for the $b$
baryons \cite{mv2}. In particular, it might seem from the large value of
$x$ that the spectator effects in the $b$ baryons can be relatively
large on the appropriate for the $b$ hadrons scale. Indeed, using the
expressions for $L_{eff}^{(3)}$ one can directly relate the splitting of
the decay rates $\Delta_b \equiv \Gamma(\Lambda_b) - \Gamma(\Xi_b^-)$ to
the differences $\Delta_1$ and $\Delta_2$ for the charmed baryons
\cite{mv2}:
\beq
\Delta_b \approx  |V_{bc}|^2  \, {m_b^2 \over m_c^2} \, (0.85 \,
\Delta_1 + 0.91 \, \Delta_2) \approx 0.015 \, \Delta_1 + 0.016 \,
\Delta_2 \approx 0.11 \pm 0.03 \, ps^{-1}~,
\label{dbres}
\eeq
which constitutes about 14\% of the total decay rate of $\Lambda_b$. The
decay rates of $\Lambda_b$ and $\Xi_b^0$ are degenerate due to the
approximation of the $U$ symmetry, used in eq.(\ref{l3nlb}). However, as
will be argued in this section, the average shift of the decay rates in
the $b$ baryon triplet due to the considered spectator effects should be
much smaller than this splitting and should amount to only of the order
of $10^{-2} \, ps^{-1}$. Thus the enhancement of the decay rate of
$\Lambda_b$ by the four-quark effects amounts to essentially one third
of the splitting in eq.(\ref{dbres}).

In order to substantiate this claim we introduce the average decay rate
in a heavy baryon triplet
\beq
{\overline \Gamma}_Q= {1 \over 3} \, \left ( \Gamma(\Lambda_Q) +
\Gamma(\Xi_Q^1)+\Gamma(\Xi_Q^2) \right )~,
\label{avgam}
\eeq
where $\Xi_Q^1$ and $\Xi_Q^2$ stand for the two heavy cascade hyperons
($\Xi_c^0$ and $\Xi_c^+$ if $Q$ is the charmed quark, and $\Xi_b^-$ and
$\Xi_b^0$ in the case of $b$ baryons). The contribution $\delta^{(3)}
{\overline \Gamma}_Q$ of the four-quark operators to ${\overline
\Gamma}_Q$ is generally expressed in terms of two flavor-singlet matrix
elements:
\begin{eqnarray}
&&x_s = {1 \over 3} \, \langle H_Q | ({\overline Q} \, \gamma_\mu \, Q)
\left ( ({\overline u} \, \gamma_\mu \, u)+({\overline d} \, \gamma_\mu
\, d)+ ({\overline s} \, \gamma_\mu \, s) \right ) | H_Q \rangle
\nonumber \\
&&y_s = {1 \over 3} \, \langle H_Q | ({\overline Q}_i \, \gamma_\mu \,
Q_k) \left ( ({\overline u}_k \, \gamma_\mu \, u_i)+({\overline d}_k \,
\gamma_\mu \, d_i)+ ({\overline s}_k \, \gamma_\mu \, s_i) \right ) |
H_Q \rangle ~,
\label{xys}
\end{eqnarray}
where $H_Q$ stands for a heavy hyperon in the (anti)triplet. Consider
the contribution of the term in eq.(\ref{l3nl}) to the dominant
nonleptonic part of the average decay rate in the charmed baryon
triplet. Using the gamma matrix Fierz identities and the color Fierz
identity $t^a_{ij} \, t^a_{kl}= \delta_{il} \, \delta_{kj}/2 - 
\delta_{ij} \, \delta_{kl}/6$, one finds for this contribution a simple
expression:
\beq
\delta_{nl}^{(3,0)} {\overline \Gamma}_c= \cos^4 \theta \, {G_F^2 \,
m_c^2 \over 8 \pi} (C_+^2 + C_-^2)\, \kappa^{5/18} \, (x_s-3 \, y_s)~.
\label{d3gc}
\eeq
Similarly, the average shift of the decay rate in the triplet of the $b$
baryons is found from eq.(\ref{l3nlb}) as
\beq
\delta^{(3)} {\overline \Gamma}_b= |V_{bc}|^2 \, {G_F^2 \, m_b^2 \over 8
\pi} ({\tilde C}_+ - {\tilde C}_-)^2\, {\tilde \kappa}^{5/18} \, (x_s-3
\, y_s)~.
\label{d3gb}
\eeq
Therefore from the later two expressions one finds that in the ratio of
the shifts the combination $(x_s-3 \, y_s)$ of the unknown hadronic
matrix elements cancels out, and that the shifts are related as
\beq
\delta^{(3)} {\overline \Gamma}_b= {|V_{bc}|^2 \over \cos^4} \,
{m_b^2 \over m_c^2} \,
{({\tilde C}_+ - {\tilde C}_-)^2 \over C_+^2 + C_-^2 } \,  \left [
{\alpha_s (m_c) \over \alpha_s(m_b)} \right ]^{5/18} \,
\delta_{nl}^{(3,0)} {\overline \Gamma}_c \approx 0.0025 \,
\delta_{nl}^{(3,0)} {\overline \Gamma}_c~.
\label{rbc}
\eeq
(One can observe, with satisfaction, that the dependence on the
unphysical parameter $\mu$ also cancels out, as it should.)
This equation shows that relatively to the charmed baryons the average
shift of the decay rates in the $b$ baryon triplet is greatly suppressed
by the ratio $({\tilde C}_+ - {\tilde C}_-)^2 / (C_+^2 + C_-^2)$, which
parametrically is of the second order in $\alpha_s$, and numerically is
only about 0.12.

An estimate of $\delta^{(3)} {\overline \Gamma}_b$ from eq.(\ref{rbc})
in absolute terms depends on evaluating the average shift
$\delta_{nl}^{(3,0)} {\overline \Gamma}_c$ for charmed baryons. The
latter shift can be conservatively bounded from above by the average
total decay rate of those baryons: $\delta_{nl}^{(3,0)} {\overline
\Gamma}_c < {\overline \Gamma_c} = 6.0 \pm 0.7 \, ps^{-1}$, which then
yields, using eq.(\ref{rbc}), an upper bound $\delta^{(3)} {\overline
\Gamma}_b < 0.015 \pm 0.002 \, ps^{-1}$. More realistically, one should
subtract from the total average width ${\overline \Gamma_c}$ the
contribution of the `parton' term, which can be estimated from the decay
rate of $D_0$ with account of the $O(m_c^{-2})$ effects \cite{buv}, as
amounting to about $3 \, ps^{-1}$. (One should also take into account
the semileptonic contribution to the total decay rates, which however is
quite small at this level of accuracy). Thus a realistic evaluation of
$\delta^{(3)} {\overline \Gamma}_b$ does not exceed $0.01 \, ps^{-1}$.

\section{Discussion}

Due to the observed here suppression of the flavor-singlet spectator
shift of the decay rates of the $b$ baryons (eq.(\ref{d3gb})), smaller
effects, which are neglected so far, may become important in this
quantity. One such effect is that of modification of the coefficients in
the expression (\ref{l3nlb}) for the spectator effects in nonleptonic
decays of $b$ baryons due to nonvanishing ratio $m_c^2/m_b^2$. By a
simple inspection of the formula with full kinematical factors in Ref.
\cite{ns} one can readily verify that the corresponding modification of
the flavor-singlet part is of order $m_c^2/m_b^2 \sim 0.1$ with no
anomalously large coefficients. Thus, the kinematical factors cannot
significantly enhance the average shift of the decay rates in the baryon
triplet. Another, more subtle, effect is that of the flavor SU(3)
symmetry breaking. If it amounts to about 30\% of the `natural' value of
the spectator effects in the $b$ hadrons, it would obviously exceed the
shift of the decay rates, given by eq.({\ref{rbc}), however it would
still be insufficiently large to explain the current data on
$\tau(\Lambda_b)$. In view of the usual difficulty of theoretically
estimating this effect, one can only rely on future data on the
difference of lifetimes of $\Lambda_b$ and $\Xi_b^0$, which would serve
as a measure of both the $O(m_c^2/m_b^2)$ corrections in the
coefficients and of the flavor SU(3) breaking in the spectator effects.
One more source of corrections to the LLO formula in eq.(\ref{rbc}) is
provided by the next-to-leading-log terms in the perturbation theory 
coefficients in the OPE. Apriori these terms are parametrically
suppressed by at least $\alpha_s(m_c)/\pi \sim 0.1$, i.e. the `natural'
magnitude of their contribution is of the same order as the LLO result
in eq.(\ref{rbc}). Whether these terms contain a large numerical
enhancement can be found out only by explicit calculation.

Thus, summarizing the reasoning of the present paper, an analysis of
spectator effects in heavy baryons, using the relations between the $b$
baryon decays and the charmed baryon decays, rather than relying on
models of baryon wave function, confirms the long-standing conclusion
that theoretically it is highly unlikely that the ratio of the lifetimes
$\tau(\Lambda_b)/\tau(B_d)$ is significantly below 0.9. Moreover, it is
found that the shift due to spectator effects of the average decay rate
of the triplet of the $b$ baryons is considerably less than the
splitting of the rates within the triplet, i.e. between the decay rate
of $\Xi_b^-$ and the (approximately) degenerate decay rates of
$\Lambda_b$ and $\Xi_b^0$. In other words, the predicted pattern of the
lifetimes of these $b$ hadrons is
\beq
\tau(\Xi_b^0) \approx \tau(\Lambda_b) < \tau(B_d) < \tau(\Xi_b^-)
\label{pattern}
\eeq
with the difference between the largest and the smallest rates given by
eq.(\ref{dbres}). The degeneracy of the decay rates of $\Lambda_b$ and
$\Xi_b^0$ is a consequence of the approximations used, in particular of
the flavor SU(3) symmetry and of neglecting the kinematical corrections
of order $m_c^2/m_b^2$. An experimental measurement of the difference of
these rates would provide a fair measure of the validity of these
approximations. \\[0.2in]

\noindent
{\large \bf Acknowledgement}

\noindent
This work is supported in part by DOE under the grant number
DE-FG02-94ER40823.

\end{document}